\documentclass[aps,pre,twocolumn,superscriptaddress,showpacs]{revtex4}
\usepackage{graphicx}
\usepackage{epsfig, amsmath, amssymb}

\newcommand{\be}{\begin{equation}}
\newcommand{\ee}{\end{equation}}
\newcommand{\beq}{\begin{eqnarray}}
\newcommand{\eeq}{\end{eqnarray}}

\begin{document}
\title{Exact explosive synchronization transitions in Kuramoto oscillators with time-delayed coupling}
\author{Hui Wu}
\affiliation{Department of Mathematics, Clark Atlanta University, Atlanta, USA}
\author{Mukesh Dhamala}
\affiliation{Department of Physics and Astronomy, Neuroscience Institute, Center for Behavioral Neuroscience, Georgia State and Georgia
  Tech Center for Advanced Brain Imaging, Georgia State University, Atlanta, USA}
\date{\today}
%Neuronal network dynamics

\begin{abstract}
  Synchronization commonly occurs in many natural and man-made systems, from neurons in the brain to cardiac cells to power grids to
  Josephson junction arrays. Transitions to or out of synchrony for coupled oscillators depend on several factors, such as individual frequencies,
  coupling, interaction time delays and network structure-function relation. Here, using a generalized Kuramoto model of
  time-delay coupled phase oscillators with frequency-weighted coupling, we study the stability of incoherent and coherent states
  and the transitions to or out of explosive (abrupt, first-order like) phase synchronization. We analytically
  derive the exact formulas for the critical coupling strengths at different time delays in both directions of increasing
  (forward) and decreasing (backward) coupling strengths. We find that time-delay does not affect the transition for the backward direction
  but shifts the transition for the forward direction of increasing coupling strength. These results provide valuable insights into our
  understanding of dynamical mechanisms for explosive synchronization in presence of often unavoidable time delays present in many
  physical and biological systems. 
\end{abstract}
\maketitle

{\it Introduction.} In many physical, biological and technological oscillatory systems, useful function emerges from collective synchronization of an ensemble of
constituent oscillators. Examples include working of neurons in the brain~\cite{Varela:2001,Engel:2001,Buzsaki:2004}, phase-locking of
Josephson junction arrays~\cite{Wiesenfeld:1996,Cawthorne:1999}, the dynamics of power grids~\cite{Motter:2013}.
The Kuramoto model~\cite{Kuramoto:1975}, originally formulated to simplify the Winfree's coupled oscillator model for the
circadian rhythms of plants and animals~\cite{Winfree:1967}, remarkably
generalizes to explain phase synchronization phenomena in these examples and many more~\cite{Strogatz:2003,Rodrigues:2016}.
Transitions to or out of synchronization as a consequence of changing coupling strength are analogous to phase transitions studied in statistical
physics such as ferromagnetic, superconductive and thermodynamic transitions~\cite{Boccaletti:2016}.

Collective synchronization of coupled phase oscillators depends on several factors: intrinsic frequency distribution, coupling strength, interaction time-delays,
network and structure (coupling strength or topology)- dynamics (frequency) relation. As the coupling strength is changed across a certain critical value,
the transition from incoherent to coherent, or coherent to incoherent states takes place smoothly (the second-order phase transition like) or
abruptly (the first-order like). When coupling is associated with oscillator characteristics or outputs, abrupt transitions to
synchrony can occur with hysteresis in a variety of coupled oscillator systems, including in Josephon junction arrays~\cite{Filtrella:2007},
in complex networks of oscillators~\cite{Pazo:2005,Gomez:2011,Peron:2012,Boccaletti:2016,Wang:2017}, and in frequency-weigthed, mean-field
coupled system of Kuramoto models~\cite{Hu:2014}. Time-delay in mean-field coupling can also make the
synchronization transition abrupt~\cite{Yeung:1999,Choi:2000}. Despite our current advanced understanding of synchronization transitions in a variety
of these systems, the effects of time-delay and frequency correlated mean-field coupling (structure-dynamics relation) in phase synchronization remain
to be explored.

Interaction time-delays and structure-function interdependence are usually unavoidable characteristics of spatially distributed,
adaptive oscillatory systems, such as neurons in the brain that has a functional organization~\cite{Dhamala:2004,Adhikari:2011,Buzsaki:2004}.
In such systems, smooth or abrupt synchronization transitions may help us to
distinguish between normal and abnormal functioning, such as pconerceptual decision-making~\cite{Adhikari:2013} as an
example of normal functions and epileptic seizure as dysfunction~\cite{Adhikari2:2013,Wang:2017}.          

In this work, we analyze a generalized Kuramoto model of time-delay coupled phase oscillators with frequency-weighted global coupling for
stability of incoherent states and coherent states and derive the exact analytical solutions for the critical coupling strengths at different
time delays in both directions of increasing (forward) and decreasing (backward) coupling strengths. Here, as a general result, we will come to show that
the time delay coupling affects the abrupt synchronization transition only in the forward direction and not in the backward direction.  

{\it Methods and results}. We consider the following generalized Kuramoto model with time-delay and frequency-weighted coupling:
\begin{eqnarray}\label{eq:0}
\dot{\theta_i}(t)= \omega_i+\frac{k}{N}|\omega_i|\sum_{j=1}^{N}\sin(\theta_j(t-\tau)-\theta_i(t)) 
\end{eqnarray}
Here, the coupled system consists of N number of oscillators, each with $\theta_i(t)$ as the instantaneous phase at time $t$, $\dot{\theta_i}(t)$ its derivative
and $\omega_i$ as the natural frequency. A set of N natural frequencies is drawn from a zero-centered symmetric distributuion
function ($g(\omega)=g(-\omega)$). The coupling strength $k$ is modulated by $|\omega_i|$.  The heterogeneity of couplings thus achieved can represent
characteristics of adaptive oscillator systems commonly found in nature.         

For a solvable analytical treatment, we consider the case of fully connected networks so that we can use the mean-field approach with the Kuramoto
order parameter $r = \frac{1}{N}|\sum_{j=1}^{N}e^{i\theta_j}|$ and the continuity equation for the time-evolution of instantaneous phase distribution
$\rho(\theta,\omega,t)$ on a unit circle. Dependent on the coupling strength $k$ and time delay $\tau$ for a given frequency distribution $g(\omega)$,
the coupled system as represented in Eq~(1), can show phase coherence ($r>0$), or incoherence ($r\approx 0$). For the forward (incoherent to coherent state)
transition, we linearize the continuity equation around the incoherent state ($\rho(\theta,\omega,t)=1/2\pi$) and obtain the critical coupling
strength ${K_f}$ for the forward direction. For the backward (coherent to incoherent state) transition, we start with fully coherent state ($r=1$) at
sufficiently large $k$ and use the self-consistency approach on the main mean-field equation to obtain the critical coupling strength ${K_b}$ in the
backward direction.  Here, we show our calculations for a zero-centered Lorentzian distribution of frequencies, but the calculation method can be
applied to any smooth symmetric frequency distribution.   

{\it Foward phase transition.} The order parameter $r$ is the ensemable average of the individual complex amplitudes:
$re^{i\phi}=\frac{1}{N}|\sum_{j=1}^{N}e^{i\theta_j}|$, where $r$ characterizes phase coherence and $\phi$ the average phase of the coupled system.   
In the continuum limit $N\to\infty$, the probability density function ($\rho(\theta,\omega,t)$) that represents the fraction of oscillators with
frequency $\omega$ whose phases are distributed between $\theta$ and $\theta+d\theta$ satisfies
%\begin{eqnarray}\label{eq:1}
(i) the normalizing condition: $\int_0^{2\pi}\rho(\theta,\omega,t)=1$ and (ii) the incoherent state value $\rho_0(\theta,\omega,t) = 1/2\pi$ uniformly
distributed over the unit circle.
%\end{eqnarray
Now, considering $r$ for $t-\tau$ time and a factor for both sides in the definition of $r$, we get
\begin{eqnarray}\label{eq:2}
r'e^{i(\phi(t-\tau)-\theta_i(t))}=\frac{1}{N}\sum_{j=1}^{N}e^{i(\theta_j(t-\tau)-\theta_i(t))}
\end{eqnarray}
The imaginary part of Eq~(\ref{eq:2}) is:
\begin{eqnarray}\label{eq:3}
r'\sin(\phi(t-\tau)-\theta_i(t))=\frac{1}{N}\sum_{j=1}^{N}\sin(\theta_j(t-\tau)-\theta_i(t))
\end{eqnarray}
With Eq~(\ref{eq:3}) and Eq~(\ref{eq:0}), we obtain:
\begin{eqnarray}\label{eq:5}
\dot{\theta_i}=\omega_i+kr'|\omega_i|\sin(\phi(t-\tau)-\theta_i(t))
\end{eqnarray}
We introduce a small perturbation to a completely incoherent state: $\rho_0(\theta,\omega,t)=\frac{1}{2\pi}$ with $\epsilon\ll 1$:
\begin{eqnarray}\label{eq:6}
\rho(\theta,\omega,t)=\frac{1}{2\pi}+\epsilon\eta(\theta,\omega,t)
\end{eqnarray}
Since
%\begin{eqnarray}\label{eq:7}
$\int_0^{2\pi}\eta(\theta,\omega,t)d\theta=0$, 
%\end{eqnarray}
we have
\begin{eqnarray}\nonumber\label{eq:8}
r'e^{i\phi}=\frac{1}{N}\sum_{j=1}^{N}e^{i\theta_j}=\int_0^{2\pi}\int_{-\infty}^{\infty}e^{i\theta}\rho(\theta,\omega,t)g(\omega)d\omega d\theta\\
=\epsilon\int_0^{2\pi}\int_{-\infty}^{\infty}e^{i\theta}\eta(\theta,\omega,t)g(\omega)d\omega d\theta=\epsilon re^{i\phi}
\end{eqnarray}
We now get $r'= \epsilon r$ and 
\begin{eqnarray}\label{eq:10}
re^{i\phi}=\int_0^{2\pi}\int_{-\infty}^{\infty}e^{i\theta}\eta(\theta,\omega,t)g(\omega)d\omega d\theta. 
\end{eqnarray}
The continuity equation for $\rho$ is:
\begin{eqnarray}\label{eq:11}
\frac{\partial\rho}{\partial t}+\frac{\partial(\rho v)}{\partial \theta}=0
\end{eqnarray}
The flow velocity function $v(t)$ is
\begin{eqnarray}\label{eq:12}
v(t)=\omega(t)+\epsilon kr|\omega(t)|\sin(\phi(t-\tau)-\theta(t))
\end{eqnarray}
By sustituting (\ref{eq:6}),(\ref{eq:10}),(\ref{eq:12}) into (\ref{eq:11}) and considering the consistency of $O(\epsilon)$ terms, we have
\begin{eqnarray}\label{eq:13}
\frac{\partial \eta}{\partial t}=-\omega\frac{\partial \eta}{\partial \theta}+\frac{k r |\omega|\cos(\phi(t-\tau)-\theta(t))}{2\pi}
\end{eqnarray}
Here $\eta(\theta,\omega,t)$ can be expanded into the following complex Fourier series:
\begin{eqnarray}\label{eq:14}
\eta(\theta,\omega,t)=c(\omega,t)e^{i\theta}+c^*(\omega,t)e^{-i\theta}+\eta^\perp(\theta,\omega,t)
\end{eqnarray}
$\eta^\perp$ represents higher Fourier harmonics terms. Now, 
\begin{eqnarray}\nonumber\label{eq:15}
re^{i(\phi(t-\tau)-\theta(t))}=e^{-i\theta(t)}re^{i\phi(t-\tau)}=\\e^{-i\theta(t)}\int_0^{2\pi}\int_{-\infty}^{\infty}e^{ix}\eta(x,\omega,t-\tau)g(\omega)d\omega dx\nonumber\\=2\pi e^{-i\theta}\int_{-\infty}^{\infty}c^*(\omega,t-\tau)g(\omega)d\omega
\end{eqnarray}
Similarly,
\begin{eqnarray}\label{eq:16}
re^{-i(\phi(t-\tau)-\theta(t))}=2\pi e^{i\theta}\int_{-\infty}^{\infty}c(\omega,t-\tau)g(\omega)d\omega
\end{eqnarray}
Now, with Eq~\ref{eq:15} and Eq~\ref{eq:16},
\begin{eqnarray}\nonumber\label{eq:17}
r\cos(\phi(t-\tau)-\theta(t))= \\ \pi[e^{-i\theta}\int_{-\infty}^{\infty}c(\omega,t-\tau)g(\omega)d\omega\nonumber\\
+e^{i\theta}\int_{-\infty}^{\infty}c^*(\omega,t-\tau)g(\omega)d\omega]
\end{eqnarray}
Using (\ref{eq:14}), (\ref{eq:17}) into (\ref{eq:13}) and comparing the coefficients of $e^{i\theta}$:
\begin{eqnarray}\label{eq:18}
\frac{\partial c(\omega,t)}{\partial t}=-i\omega c(\omega,t)+\frac{k |\omega|}{2}\int_{-\infty}^{\infty}c(v,t-\tau)g(v)dv
\end{eqnarray}
We now look for a separable solution: $c(\omega,t)=b(\omega)e^{\lambda t}$ in the equation (\ref{eq:18}):
\begin{eqnarray}\label{eq:19}
\lambda b(\omega)=-i\omega b(\omega)+\frac{k |\omega|e^{-\lambda\tau}}{2}\int_{-\infty}^{\infty}b(v)g(v)dv
\end{eqnarray}
If we let $\frac{k|\omega|}{2}\int_{-\infty}^{\infty}b(v)g(v)dv=A$, then from (\ref{eq:19}), $b(\omega)e^{\lambda\tau}=\frac{|\omega|A}{\lambda+i\omega}$,
we have from above equations, A is cancelled:

\begin{eqnarray}\label{eq:37}
e^{\lambda\tau}=\frac{k}{2}\int_{-\infty}^{\infty}\frac{\lambda |\omega|}{\lambda^2+\omega^2}g(\omega)d\omega
\end{eqnarray}
When natural frequencies follow a standard Lorentzian distribution,  $g(\omega)=\frac{1}{\pi(\omega^2+1)}$, 
\begin{eqnarray}\label{eq:39}
k=\frac{\pi(\lambda^2-1)e^{\lambda\tau}}{\lambda In\lambda}
\end{eqnarray}
$R=|\lambda|>0$, $k$ passes through the bifurcation point when $\lambda=iR$ or $\lambda=-iR$.
When $\lambda=iR$,
\begin{eqnarray}\label{eq:40}
k_1=\frac{\pi(-R^2-1)e^{iR\tau}}{iR(In R+\frac{\pi}{2}i)}
\end{eqnarray}
When $\lambda=-iR$,
\begin{eqnarray}\label{eq:41}
k_2=\frac{\pi(-R^2-1)e^{-iR\tau}}{-iR(In R-\frac{\pi}{2}i)}
\end{eqnarray}
We set $k_1=\bar{k_2}$ (conjugate pairs) to seek for real critical values. When $\tau$ and $R$ satisfies:
\begin{eqnarray}\label{eq:43}
\tan(\tau R)=-\frac{2}{\pi}In R
\end{eqnarray}
$k_1=k_2$ and both are real. Equation (\ref{eq:43}) has many number of intersection points.
Only one $R=R_0>0$ is a unique efficient solution and all others are extra roots.
The forward critical value of $k$ is determined by the value of $R_0$ as follows:
\begin{eqnarray}\label{eq:46}
K_f=\frac{2(R_0^2+1)}{R_0}\cos(\tau R_0)
\end{eqnarray}
All unique solutions in (\ref{eq:46}) are greater than or equal to 4 for any $\tau\ge 0$), as shown in Fig~1 (a). 
%%%%%%%%%%%%%%%%%%%%%%%%%%%%%%%%%%%%%%%%%%%%
\begin{figure}
\epsfig{figure=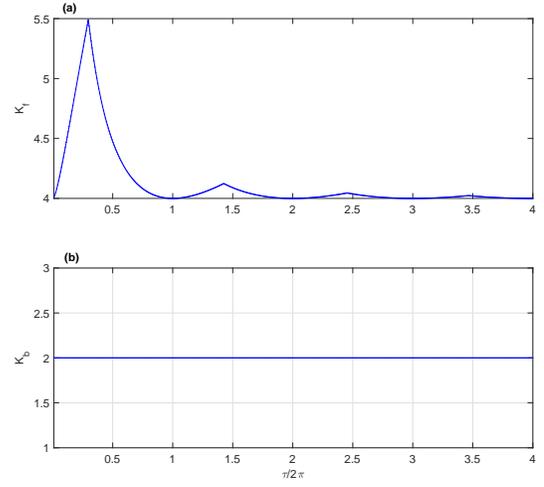,width=0.8\linewidth}
\caption{(a, b). Theoretical predictions of critical coupling strengths ($K_{critical}=\{K_f,K_b\}$) as a function of time delay ($\tau$) for
  increasing (forward) and decreasing (backward) directions.}
\label{fig:fig1}
\end{figure}
%%%%%%%%%%%%%%%%%%%%%%%%%%%%%%%%%%%%%%%%%%%%

{\it Backward phase transition.} In the case of fully connected network, we have
\begin{eqnarray}\label{eq:49}
\frac{1}{N}\sum_{j=1}^N e^{i(\theta_j(t-\tau)-\theta_i(t))}=r e^{i(\phi(t-\tau)-\theta_i(t))}
\end{eqnarray}
Considering only the imaginary part of (\ref{eq:49}), we have
\begin{eqnarray}\label{eq:50}
\frac{1}{N}\sum_{j=1}^N \sin(\theta_j(t-\tau)-\theta_i(t))=r \sin(\phi(t-\tau)-\theta_i(t))
\end{eqnarray}
Substituting into (\ref{eq:0}), we get the mean-field equation:
\begin{eqnarray}\label{eq:51}
\dot{\theta_i}(t)= \omega_i(t)+kr|\omega_i(t)|\sin(\phi(t-\tau)-\theta_i(t))
\end{eqnarray}
Set a rotating frame with the average phase of the system,
\begin{eqnarray}\label{eq:52}
\phi(t)=\phi(0)+\langle \omega\rangle t
\end{eqnarray}
Here $\langle \omega\rangle$ is the average frequency of the oscillators. For a symmetric distribution of $g(\omega)$, $\langle \omega\rangle=0$.
Hence, $\phi(t)=\phi(0)$, $\phi(t-\tau)=\phi(0)$.
With $\Delta\theta_i(t)=\theta_i(t)-\phi(0)$, then $\Delta\theta_i(t)=\theta_i(t)-\phi(t-\tau)$, the mean field equation (\ref{eq:51}) can be transferred into:
\begin{eqnarray}\label{eq:53}
\Delta\dot{\theta_i}=\omega_i-kr|\omega_i|\sin(\Delta\theta_i)
\end{eqnarray}
In the coherent state, all the oscillators are phase locked. So $\Delta\dot{\theta_i}=0$.
\begin{eqnarray}\label{eq:54}
\Delta\theta_i = \left\{
        \begin{array}{ll}
            arc\sin(\frac{1}{kr}) & \quad \omega_i > 0 \\
            arc\sin(-\frac{1}{kr}) & \quad \omega_i< 0
        \end{array}
    \right.
\end{eqnarray}
\begin{eqnarray}\label{eq:55}
r=\frac{1}{N}\sum_{j=1}^N e^{i\Delta\theta_j}=\frac{1}{2}(e^{i\Delta\theta_{i+}}+e^{i\Delta\theta_{i-}})
\end{eqnarray}
$\Delta\theta_{i+}$ and $\Delta\theta_{i-}$ represents the two groups in the equation (\ref{eq:54}).
\begin{eqnarray}\label{eq:56}
r=\frac{1}{2}(\cos(\Delta\theta_{i+})+\cos(\Delta\theta_{i-}))
\end{eqnarray}
$\sin(\Delta\theta_{i+})=\frac{1}{kr}$, $\sin(\Delta\theta_{i-})=-\frac{1}{kr}$ and $r=\cos(\Delta\theta_{i+})=\cos(\Delta\theta_{i-})=\sqrt{1-(\frac{1}{kr})^2}$
Hence,
\begin{eqnarray}\label{eq:57}
r^2=\sqrt{r^2-\frac{1}{k^2}}.
\end{eqnarray}
%%%%%%%%%%%%%%%%%%%%%%%%%%%%%%%%%%%%%%%%%%%%
\begin{figure}
\epsfig{figure=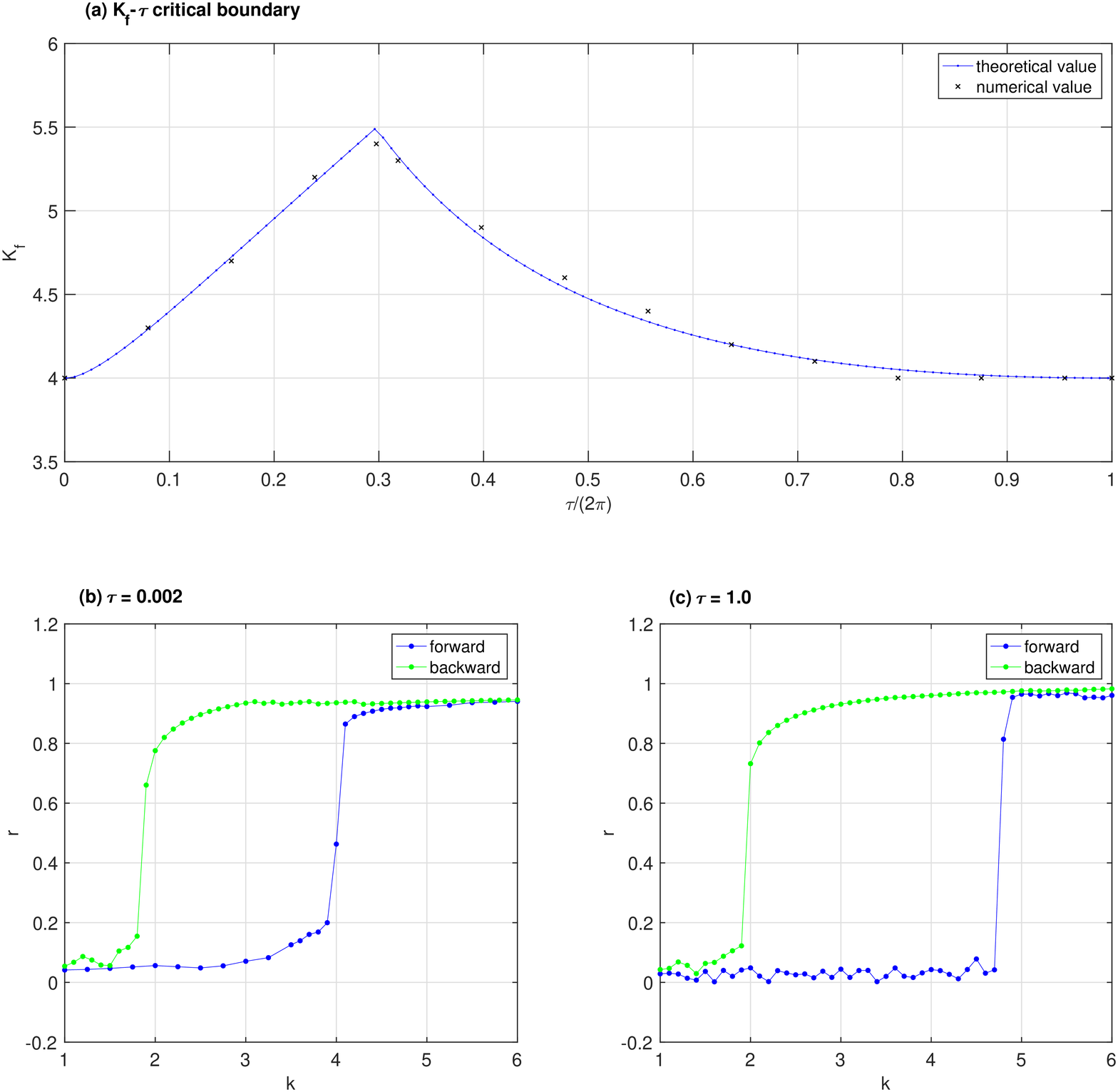,width=0.9\linewidth}
\caption{a. Numerical values (x) overlaid on the theoretically predicted critical values for the forward transition ($K_f-\tau$ boundary),
  (b) $r$ versus $k$ at $\tau/2\pi = 3.2\times 10^{-4}$, and
  (c) $r$ versus $k$ at $\tau/2\pi = 0.16$. These numerical results are based on the RK4-integration scheme with
  step-size $ = 0.001$ to solve the ordinary differential equations for $1000$ coupled oscillators with incoherent initial conditions for
  the forward direction.}
\label{fig:fig2}
\end{figure}
%%%%%%%%%%%%%%%%%%%%%%%%%%%%%%%%%%%%%%%%%%%%
The above equation (\ref{eq:57}) has a solution for $r>0$ iff $k\ge 2$.
Thus, the backward critical value of $k$ becomes $K_b = 2$ for any even, symmetric distribution function $g(\omega)$ (Fig~1 (b)).
As shown in Fig~2, we also numerically verify these analytical results.

{\it Conclusions}. Here, we generalize the Kuramoto model of globally coupled phase oscillators
with time-delay and oscillation frequency-modulated coupling considering its relevance to adaptive physical,
biological or technoligcal oscillators. We have analytically and numerically studied the stability of first-order synchronization
in this generalized Kuramoto model. We have found the exact formulas for the critical coupling strengths at different time delays in both
the increasing (forward) and decreasing (backward) directions of coupling strengths. We find that time-delay does not change the transition
in the backward direction but shifts the transition for the forward direction. These results provide useful insights into our
understanding of dynamical mechanisms leading to explosive synchronization in presence of often unavoidable
time delays in spatially distributed and functionally organized systems. We envision that our theoretical work may encourage future research
on abrupt collective synchronization in models of spatially distributed and functionally organized
real systems that can be mapped or reduced onto the Kuramoto model, such as Josehpon juctions~\cite{Wiesenfeld:1996},
cortical neurons~\cite{Sadilek:2015} and many more~\cite{Hu:2014}.

\end{document}